\documentstyle[12pt]{article}
\newcommand{\be}{\begin{equation}}
\newcommand{\ee}{\end{equation}}
\newcommand{\bea}{\begin{eqnarray}}
\newcommand{\eea}{\end{eqnarray}}
\newcommand{\nn}{\nonumber}
\newcommand{\p}{\phi}
\newcommand{\q}{\omega}
\newcommand{\hp}{\hat{\phi}}
\newcommand{\hq}{\hat{\omega}}
\newcommand{\tp}{\tilde{\phi}}
\newcommand{\tq}{\tilde{\omega}}
\newcommand{\op}{\hat{P}}

\newcommand{\F}{\cal{F}}

\begin{document}
\title{Abelian Subset of Second Class Constraints}
\author{F. Loran\thanks{e-mail:
loran@cc.iut.ac.ir}\\ \\
  {\it Department of  Physics, Isfahan University of Technology (IUT)}\\
{\it Isfahan,  Iran,} \\{\it and}\\
  {\it Institute for Studies in Theoretical Physics and Mathematics (IPM)}\\
{\it P. O. Box: 19395-5531, Tehran, Iran.}}
\date{}
\maketitle
\begin{abstract}
We show that after mapping each element of a set of second class constraints to the
surface of the other ones, half of them form a subset of abelian first class
constraints. The explicit form of the map is obtained considering the most general
Poisson structure. We also introduce a proper redefinition of second class
constraints that makes their algebra symplectic.
\end{abstract}
\newpage
\section{introduction}
When Dirac introduced constrained systems \cite{Dirac}, he classified constraints as
first class and second class. First class constraints have been interesting since they
turned out to be generators of gauge transformation. These constraints introduced a
new class of symmetries, which for example, lead to Ward identities in the context of
renormalization \cite{Henbook}. The main requirement in quantization of first class
constraints is the covariance of observables under gauge transformations. In Dirac
quantization, this requirement is satisfied by considering physical states as null
eigen states of the generator of gauge transformation. The same idea is followed in
BRST where a nilpotent BRST-charge generates BRST-transformation \cite{Hen}.
 \par
There are two major difficulties in both Dirac quantization and BRST. In general,
first class constraints satisfy a closed algebra in which the structure coefficients
are some functions of phase space coordinates. Consistency of these methods of
quantization depends on the possibility of a definite operator ordering; the structure
coefficients should stand on the left side of first class constraints. Another
problem is obtaining the explicit form of the generator of gauge transformation or
BRST-charge. Both difficulties can be overcome by making the first class constraints
abelian \cite{Abelian}.
\par
Second class constraints were thought to be redundant degrees of freedom that one
should get rid of them before quantization, for example by using Dirac bracket instead
of Poisson bracket. But second class constraints are more important. For example, in
reference \cite{Under}, the gauge theory of second class systems is discussed. Or in
closed string theory, it is claimed that boundary conditions lead to a set of second
class constraints which give rise to non-commutativity of space-time \cite{String}.
On the other hand, covariant quantization, in general, is not consistent with
classification of constraints as first and second class \cite{Covariant}.
Consequently, we need a general method of quantization which treats both classes on
the same footing.
 \begin{enumerate}
 \item One possibility is to convert second class constraints to first class \cite{Convert}.
 Given a constraint system
possessing second class constraints, in principle, one can consider an extended phase
space and redefine second class constraints and the Hamiltonian to find an equivalent
first class system. There are two difficulties in doing so. Firstly, it is not so
easy to find out such redefinitions in general cases. Secondly, assuming that the
conversion is done, one may still encounter the above mentioned difficulties in
quantization of first class constraints. As is well known, all these problems can be
remedied most easily provided one makes the algebra of second class constraints
symplectic.
\item  Another possibility is to consider half of second class constraints
 as first class constraints and the remaining ones as gauge fixing conditions
 \cite{Henbook}. This method, for example, is used to study gauge invariance in the
 Proca model \cite{Vyth}.
 \end{enumerate}
 In this paper, we prove that after mapping each element of a set of second class constraints
 to the surface of the other ones, half of them form an {\it abelian} subset.
 In addition we present a general method for redefining second class
constraints to make their algebra symplectic. Although this method may not preserve
covariance but it is still interesting since it works globally, and for a general
Poisson structure. Therefore it provides a simple conversion of second class
constraints to first class ones.
 \par
 In reference \cite{Abelian} it is shown that
first class constraints become abelian when they are mapped to the surface of each
other. Thus, it seems that in this way, one can obtain abelian subset of a given set
of constraints in the most simple way.
\par
The organization of paper is as follows. In section 2, we introduce necessary
definitions and lemmas. The method is introduced in section 3. We conclude our
results in section 4.
\section{Definitions and Lemmas}
In this section we provide some general tools necessary for arguments of the next
section. Consider a phase space defined by a set of coordinates $z^\mu$ satisfying the
Poisson algebra,
 \be
 \{z^\mu,z^\nu\}=J^{\mu\nu}(z),
 \label{a1}\ee
 in which $J^{\mu\nu}(z)$ is a full rank anti-symmetric tensor, e.g. the symplectic two form:
 \be
 J=\left(
 \begin{array}{cc}
 0&+1\\
 -1&0
 \end{array}
 \right)
 \label{a2} \ee
 Assume a pair of conjugate functions $\phi(z)$ and $\q(z)$ in $\F$, satisfying the relation,
 \be
 \{\p,\q\}=1,
 \label{a3}
 \ee
 where $\F$ stands for the set of real analytic functions of the phase space
 coordinates. In fact for a given $\p \in \F$, using the Cauchy-Kowalevski theorem \cite{John},
 one can show that there exist at least one function $\q\in \F$ that satisfies the
 relation,
 \bea
 \{\p,\q\}&=&a^\mu(z)\frac{\partial \q}{\partial z^\mu}=1,\nn\\
 a^\mu&=&\frac{\partial \p}{\partial z^\nu}J^{\nu\mu}.
 \label{a4}
 \eea
% At the end of this section we introduce a method to obtain such $\q$'s, explicitly.
 Corresponding to each $\xi\in \F$ an operator $\hat{\xi}:\F \to \F$  can be defined as
 follows,
 \be
 \hat{\xi}\chi=\{\xi,\chi\},\hspace{1cm}\chi\in\F.
 \label{a5}\ee
 It is easy to verify that,
 \bea
 \hat{\xi}\left(\chi_1\chi_2\right)&=&\left\{\xi,(\chi_1\chi_2)\right\}\nn\\
 &=&\chi_1\{\xi,\chi_2\}+\{\xi,\chi_1\}\chi_2\nn\\
 &=&\left(\hat{\xi}\chi_1\right)\chi_2+
 \chi_1\left(\hat{\xi}\chi_2\right).
 \label{a6}
 \eea
 Considering the operators $(\hp,\hq)$ where $\{\p,\q\}=1$, from Eq.(\ref{a3}) one can show that these operators
 satisfy the following relations:
 \bea
 \label{a7-1}
 &&[\hp,\hq]=0,\\
 \label{a7-2}
 &&[\hp,\p]=[\hq,\q]=0,\\
 \label{a7-3}
 &&[\hp,\q]=[\p,\hq]=1.
 \eea
 These properties can be easily verified. For example,
 for an arbitrary function $\xi\in\F$, we have,
 \bea
 [\hp,\hq]\xi&=&\left\{\p,\{\q,\xi\}\right\}-\left\{\q,\{\p,\xi\}\right\}\nn\\
 &=& -\left\{\xi,\{\p,\q\}\right\}\nn\\
 &=&-\{\xi,1\}\nn\\
 &=&0,
 \label{a8}
 \eea
 where in the second equality we have used the Jaccobi identity.
 Considering the operators $\op_\p$ and $\op_\q$ \cite{Vyth},
 \bea
 \op_\p&\equiv&\sum_{n=0}\frac{1}{n!}\p^n\hq^n,\nn\\
 \op_\q&\equiv&\sum_{n=0}\frac{(-1)^n}{n!}\q^n\hp^n,
 \label{a9}\eea
 one can use Eqs.(\ref{a7-1},\ref{a7-2}) to show that $[\hp,\op_\p]=[\hq,\op_\q]=0$,
 and consequently,
 \be
 [\op_\p,\op_\q]=0.
 \label{a11}
 \ee
 \par
 {\bf Lemma 1}. The operators $\op_\p$ and $\op_\q$ satisfy the following
 properties:
 \bea
 \label{a12-1}
 \hq \op_\p&=&0,\\
 \label{a12-2}
 \hp \op_\q&=&0,
 \eea
 \par
 {\bf Proof.} We proof the first equality. The second equality can be proved in the
 same way. Using Eq.(\ref{a7-3}), one can show that $[\hq,\p^n]=-n\p^{n-1}$. Thus,
 \bea
 \hq\op_\p&=&[\hq,\op_\p]+\op_\p\hq\nn\\
 &=&\sum_{n=0}\frac{1}{n!}[\hq,\p^n]\hq^n+\op_\p\hq\nn\\
 &=&-\sum_{n=1}\frac{1}{(n-1)!}\p^{n-1}\hq^n+\op_\p\hq\nn\\
 &=&0.
 \label{a13}
 \eea
 \par
 {\bf Lemma 2.} Given conjugate functions $\q,\p \in\cal{F}$, the operator $\op_\p$ is
 the projection map to the subspace of the phase space defined by $\p=0$.
 \par
 {\bf Proof.}
 Using Eq.(\ref{a12-1}), it can be shown that $\op_\p^2=\op_\p$.
 Assuming the (canonical) coordinate transformation,
 \be
 z^\mu \to \p,\q, Z^{\mu'},
 \label{a14-2}
 \ee
 one verifies that $\hq=-\frac{\partial}{\partial \p}$. Therefore, for an arbitrary
 $\xi\in\F$,
 \bea
 \op_\p\xi(\p,\q,Z^{\mu'})&=&
 \sum_{n=0}\frac{(-1)^n}{n!}\p^n\frac{\partial^n}{{\partial \p}^n}\xi(\p,\q,Z^{\mu'})\nn\\
 &=&\xi(0,\q,Z^{\mu'})\nn\\
 &=&\xi|_\p.
 \label{a15}
 \eea
 \par
 {\bf Corollary 1:} The operator $\op_\q$ is projection map to the subspace $\q=0$.\\
 This corollary can be proved noting that $\hp=\frac{\partial}{\partial \q}$ and
 consequently $\op_\q\xi=\xi(\p,0,Z^{\mu'})$.
 \par
 {\bf Lemma 3.} The operator $\op\equiv\op_\p\op_\q$ is the
 projection map to the subspace, $\p=\q=0$.
 \par
 {\bf Proof.} From Eqs.(\ref{a11}-\ref{a12-2}), one obtains $\hp\op=\hq\op=\op$, thus
 $\op^2=\op$.
 Reviewing the proof of lemma 2, one verifies that
 \bea
 \op\xi(\p,\q,Z^{\mu'})&=&\op_\q\left(\op_\p\xi(\p,\q,Z^{\mu'})\right)\nn\\
 &=&\op_\q\xi(0,\q,Z^{\mu'})\nn\\
 &=&\xi(0,0,Z^{\mu'}).
 \eea
\par
{\bf Corollary 2.} If $\xi=\xi|_{\p,\q}$ then $\{\p,\xi\}=\{\q,\xi\}=0$. \\
The second equality, for example, can be proved noting that $\xi|_{\p,\q}=\op\xi$ and
$\hq\op=0$.
 \par
 The above results become practically interesting if conjugate to a given
 $\p\in\cal{F}$,
 one can obtain explicitly a function $\q$ that satisfies Eq.(\ref{a3}).
 This can be easily done if there exist a function $H\in\cal{F}$,
 such that $\hp H\neq 0$ but $\hp^{M+1} H=0$,
 for an integer $M\geq 1$. Since, in principle, $\q$ exists and satisfies Eq.(\ref{a3}),
 using the coordinate transformation (\ref{a14-2}),
 $H$ can be written as a polynomial in $\q$,
 \bea
 H\left(z(\q,\p,Z^{\mu'})\right)&=&\sum_{m=0}^M \frac{A_m(0,\p,Z^{\mu'})}{m!}\q^m,\nn\\
 \hp A_m&=&\frac{\partial}{\partial \q}A_m(0,\p,Z^{\mu'})=0.
 \label{a16}
 \eea
 Thus,
 \bea
 \label{a17-1}
  \hp^M H&=&\frac{\partial^M}{{\partial \q}^M}H=A_M,\\
  \label{a17-2}
 \hp^{M-1}H&=&A_M\q+A_{M-1}.
 \eea
 Comparing Eq.(\ref{a17-1}) with Eq.(\ref{a17-2}), one can verify $\q$ as the
 coefficient of $\hp^MH$ in $\hp^{M-1}H$.
 \par
 As an example suppose,
 \bea
 \p&=&e^x-1,\nn\\
 H&=&\frac{1}{2}p_x^2.
 \eea
 A simple calculation shows that $M=2$ and
 \bea
 \label{ex1}
 \hp H&=&e^x p_x,\\
 \label{ex2}
 \hp^2 H&=&e^{2x}.
 \eea
Comparing Eq.(\ref{ex1}) with Eq.(\ref{ex2}) one reads $\q=e^{-x}p_x$. This method can
be used to obtain gauge fixing conditions conjugate to first class constraints
\cite{Abelian}.
\par
 {\bf Lemma 4.} Considering a function $\xi\in\F$ and a conjugate pair of functions
 $\p$ and $\q$, we have $\xi=\xi|_{\p}$ iff $\hq \xi=0$.
 \par
 {\bf Proof.}
 \par
 $a)$ If $\xi=\xi|_{\p}$ then from lemma 2, $\xi=\op_\p \xi$. Therefore using Eq.(\ref{a12-1}),
 $\hq \xi=\hq \op_\p \xi=0$.
 \par
 $b)$ if $\hq \xi=0$ then $\xi=\op_\p \xi=\xi|_{\p}$.
 \par
 {\bf Corollary 3.} For arbitrary functions $\xi$ and $\chi$ in $\F$,
 \bea
  \label{aa6}
 \{\xi|_{\p},\zeta|_{\p}\}&=&\{\xi|_{\p},\zeta|_{\p}\}|_\p,\\
 \{\xi|_{\p},\p\}&=&\{\xi|_{\p},\p\}|_\p.
 \label{aa7}
 \eea
 Corollary 3 can be proved using the Jaccobi identity to show that the Poisson
 brackets of the LHS of Eqs.(\ref{aa6},\ref{aa7}) with $\q$ is vanishing.
 \par
 {\bf Lemma 5.} If $\p=\p|_{\psi}$ then $\psi=\psi|_{\p}$.
 \\
 {\bf Proof.} Since there exist a function $\q$ conjugate to $\p$,
 one can write $\psi$ as a polynomial in $\p$
 (similar to Eq.(\ref{a16})),
 \be
 \psi=\sum_{i=1}a_i\p^i+\psi|_{\p},
 \ee
 where $\hq a_i=0$, $i\geq1$. If $a_i$'s do not vanish, the assumption $\p=\p|_{\psi}$
 implies that $\psi(\p)=0$. Thus if $\psi\neq0$ then $a_i$'s should vanish and
 $\psi=\psi|_{\p}$.
 \par
{\bf Lemma 6.} If $\q_1$ and $\q_2$ are conjugate to $\p_1$ and $\p_2$ respectively,
 and
 \bea
 \label{la1}
 \p_2=\p_2|_{\p_1},\\
 \label{la2}
 \q_2=\q_2|_{\p_1},
 \eea
 then the operators $\op_{\p_1}$ and $\op_{\p_2}$ commute with each other.
 \par
 {\bf Proof.} It is sufficient to prove that $[(\p_1^n\hq_1^n),(\p_2^m\hq_2^m)]=0$.
 Using the Jaccobi identity and lemma 4, one can show that,
 \bea
 &[\hq_1,\hq_2]=\{\q_1,\q_2\}=0,&\nn\\
 &[\hq_1,\p_2]=\{\q_1,\p_2\}=0.&
 \eea
 From lemma 5 and Eq.(\ref{la1}) one verifies that $\p_1=\p_1|_{\p_2}$,
 thus $[\p_1,\hq_2]=\{\p_1,\q_2\}=0$. This completes the proof.
 \par
 {\bf Corollary 4.} The operators $\op_{\p_1}$ and $\op_{\q_2}$ commute i.e.
 $[\op_{\p_1},\op_{\q_2}]=0$.
 \par
 {\bf Corollary 5.} The operators $\op_i=\op_{\q_i}\op_{\p_i}$, $i=1,2$, commute if
 $\q_2=\op_1\q_2$ and $\p_2=\op_1\p_1$.\\
 This can be proved using lemma 3 and corollary 2.
 \par
 {\bf Lemma 7.} If $\p=\p|_{\psi}$, $\{\p,\psi\}=0$ and $\{\p,\q\}=1$, then
 $\{\p,\q|_{\psi}\}=1$.
 \par {\bf Proof.}
 Writing $\q$ as a polynomial in $\psi$,
 \be
 \q=\sum_{i=1}a_i\psi^i+\q|_{\psi},
 \ee
 one verifies that,
 \be
 1=\{\p,\q\}=\sum_{i=1}\{\p,a_i\}\psi^i+\{\p,\q|_{\psi}\}.
 \ee
 Thus,
 \be
 1=\{\p,\q|_{\psi}\}_{\psi}=\{\p,\q|_{\psi}\}_\psi=\{\p,\q|_{\psi}\},
 \ee
 where in the third equality we have used Eq.(\ref{aa6}).
 \par
 {\bf Corollary 6.} If $\{\p,\q\}=1$, then $\{\p,\q|_\p\}=1$.\\
 Using Eq.(\ref{aa7}), the proof is similar to the proof of lemma 7.
 \par
 {\bf Lemma 8.} If $\xi=\xi|_\p$, $\psi=\psi|_\p$ and $\{\p,\psi\}=0$, then
 $\tilde{\xi}=\tilde{\xi}|_\p$ in which $\tilde{\xi}\equiv\xi|_\psi$.
 \par
 {\bf Proof.} Lemma 7 implies that there exist a function $\q_\psi$ conjugate to
 $\psi$ such that $\q_\psi=\q_\psi|_\p$. Consequently from lemma 6, we know that
 $[\op_\psi,\op_\p]=0$. In addition, $\tilde{\xi}=\op_\psi\xi$
 and $\xi=\op_\p\xi$ (see lemma 2). Thus,
 \be
 \tilde{\xi}=\op_\psi\op_\p\xi=\op_\p\op_\psi\xi=\op_\p\tilde{\xi}.
 \ee
 This completes the proof.
 \section{Redefinition of Second Class Constraints}
 In this section we show that the subspace $\cal{M}$ of the phase space, defined by a
set of irreducible second class constraints,
 \be\p_a=0, \hspace{1cm}a=1,\cdots,2k,
 \label{start}\ee
  which satisfy the relation,
 \be
 \det\left(\{\p_a,\p_b\}\right)_{\cal{M}}\neq 0,
 \label{det}\ee
can be equivalently determined by a set of constraints $\tp_i$, $\tq_i$,
$i=1,\cdots,k$, satisfying the symplectic algebra,
 \bea
 \{\tp_i,\tp_j\}=0,\nn\\
 \{\tp_i,\tq_j\}=\delta_{ij},\nn\\
 \{\tq_i,\tq_j\}=0.
 \label{res}
 \eea
 For this reason, we consider the following lemmas.
 \par
 {\bf Lemma 9.} There exist at least one constraint, say $\p_{k+1}$, such that
 \be
 \{\p_1,\p_{k+1}\}_{\cal{M}}\neq 0.
 \label{f2}
 \ee
 \par
 {\bf Proof.} If it was not the case, i.e. if $\{\p_1,\p_a\}_{\cal{M}}= 0$, $a=1,\cdots,2k$,
 then,
 \be
 \det\left(\{\chi_a,\chi_b\}\right)_{\cal{M}}=0,
 \ee
 contrary to the assumption Eq.(\ref{det}).\\
 Consider the constraints $\p_1$ and $\p_{k+1}$
 and the definition,
 \be
 \label{h00}
 \q'_1\equiv \q_1-\q_1|_{\p_{k+1}},
 \ee
 where $\q_1\in \F$ is some function conjugate to $\p_1$.
 \par
{\bf Lemma 10.} If the equation $\p_{k+1}=0$ has a unique solution (the uniqueness
condition) then the constraint $\q'_1\approx 0$ is equivalent to $\p_{k+1}$.
\par
 {\bf Proof.} Using the uniqueness condition, we show that $\p_{k+1}=0$ iff
 $\q'_1=0$.\\
 Consider the coordinate transformation $z^\mu\to(\q_1,\p_1,Z^{\mu'})$.
 The assumption,
 \be
 \{\p_1,\p_{k+1}\}=\frac{\partial}{\partial \q_1}\p_{k+1}\neq 0,
 \label{le9-1}
 \ee
 reads,
 \be
 \p_{k+1}=\q_1\chi(\q_1,\p_1,z')+\xi(0,\p_1,Z^{\mu'}),
 \label{le9-2}
 \ee
 for some functions $\chi$ and $\xi$.
 From Eq.(\ref{le9-2}) one can determine $\q^0_1\equiv\q_1|_{\p_{k+1}}$ as the solution of
 equation,
 \be
 \q^0_1\chi(\q^0_1,\p_1,Z^{\mu'})+\xi(0,\p_1,z')=0.
 \ee
 Inserting $\xi$ from the above relation in Eq.(\ref{le9-2}), one verifies that,
 \bea
 \p_{k+1}&=&\q_1\chi(\q_1,\p_1,Z^{\mu'})-\q^0_1\chi(\q^0_1,\p_1,Z^{\mu'})\nn\\
 &=&\q_1\left(\chi(\q^0_1,\p_1,Z^{\mu'})+(\q_1-\q^0_1)\chi'(\q_1,\p_1,Z^{\mu'})\right)-
 \q^0_1\chi(\q^0_1,\p_1,Z^{\mu'})\nn\\
 &=&(\q_1-\q^0_1)\left(\chi(\q^0_1,\p_1,Z^{\mu'})+\q_1\chi'(\q_1,\p_1,Z^{\mu'})\right)\nn\\
 &=&\q'_1\tilde{\chi},
 \label{le9-3}
 \eea
 where $\chi'$, in the second equality, is some function that can be determined
 in terms of $\chi$ using Taylor expansion. In the last equality we have used
 definition (\ref{h00}). From Eq.(\ref{le9-3}) one finds two possible solutions
 for equation $\p_{k+1}=0$; ${\q'}^1_1=0$ and/or $\tilde{\chi}=0$. Due to uniqueness
 condition these two solutions, if both possible, should coincide. Therefore
 $\p_{k+1}=0$ if and only if ${\q'}^1_1=0$.
 Of course $\tilde{\chi}$ is non vanishing because,
 \bea
 \det\left(\{\p_a,\p_b\}\right)_{\cal{M}}&=&\pm\det
 \left(\begin{array}{ccc}
 0&\{\p_1,\p_{k+1}\}&\cdots\\
 \{\p_{k+1},\p_1\}&0&\cdots\\
 \vdots&\vdots&\vdots
 \end{array}\right)_{\cal{M}}\nn\\&=&\pm {\tilde{\chi}}^2\det
 \left(\begin{array}{ccc}
 0&\{\p_1,\q'_1\}&\cdots\\
 \{\q'_1,\p_1\}&0&\cdots\\
 \vdots&\vdots&\vdots
 \end{array}\right)_{\cal{M}}\neq0.
 \eea
 The above equation implies that not only the constraint $\q'_1$ is equivalent to
 $\p_{k+1}$ but also the set of constraints $\p_a$
 in which $\p_{k+1}$ is replaced by $\q'_1$ are second class.
 \par
 {\bf Lemma 11.} The function $\q'_1$ is conjugate to $\p_1$, i.e. $\{\p_1,\q'_1\}=1$.
 \par
 {\bf Proof.} If $\q^0_1=0$, then proof is trivial.
 If $\q^0_1\neq 0$, one can prove lemma 11 as follows.
 Consider the Taylor expansion of $\p_{k+1}$ in terms of $\q_1$,
 \be
 \p_{k+1}(\q_1,\p_1,Z^{\mu'})=\sum_{m=0}A_m(0,\p_1,Z^{\mu'})\q_1^m.
 \ee
 Since $\q^0_1=\q_1|_{\p_{k+1}}$, we have,
 \be
 \sum_{m=0}A_m(0,\p_1,Z^{\mu'})(\q_1^0)^m=0.
 \label{h3}\ee
 Consequently,
 \be
 \{\p_1,\q_1^0\}\sum_{m=1}mA_m(\q_1^0)^{m-1}=0.
 \label{h4}\ee
 This has two solutions:\par
 1)  $A_{m>0}=0$. In this case, the Poisson bracket of
  $\p_{k+1}=A_0(0,\p_1,Z^{\mu'})$ and $\p_1$ vanishes contrary to the assumption Eq.(\ref{f2}).
 \par
 2) $\{\p_1,\q^0_1\}=0$, which is the desired result.
  \par
  Let's define $\tp_1\equiv\p_1$ and $\tq_1\equiv\q'_1$.
  Using lemma 3 and corollary 2, one can make the Poisson
bracket of $\tilde{\p}_1$ and $\tilde{\p}_{k+1}$ with the other constraints vanishing
by redefining the constraints $\p_i$ and $\p_{k+i}$ ($i>1$) as follows,
 \bea
 \p_i& \to &\op_1\ \p_i,\hspace{1cm}i=2\cdots,k,\nn\\
 \p_{i+k}& \to &\op_1\ \ \p_{i+k},
 \label{red}
 \eea
 where $\op_1=\op_{\tp_1}\op_{\tq_1}$.
 Let us call these new constraints $\p^1_{a_1}$, $a_1=1,\cdots,2k^1$,
 where $k^1=k-1$.
 The determinant of the matrix of Poisson brackets of the second class constraints
 $\tp_1$, $\tq_1$ and $\p^1_{a_1}$'s is,
 \be
 \det
 \left(\begin{array}{cccc}
 0&+1&0\cdots0\\
 -1&0&0\cdots0\\
 0&0&&\\
 \vdots&\vdots&\left(\{\p^1_{a_1},\p^1_{b_1}\}\right)&\\
 0&0&&
 \end{array}\right)=\det\left(\{\p^1_{a_1},\p^1_{b_1}\}\right)_{\cal{M}}\neq0.
 \label{sym1-1}
 \ee
 Consequently there exist a constraint, say $\p^1_{k^1+1}$,
 such that $\{\p^1_1,\p^1_{k^1+1}\}_{\cal{M}}\neq 0$. From corollary 2 we know
 that $\{\p^1_1,\tp_1\}=\{\p^1_1,\tq_1\}=0$. Thus, lemma 7 guarantees
 the existence of a function $\q^1_1$ conjugate to $\p^1_1$ such that
 $\q^1_1=\q^1_1|_{\tp_1,\tq_1}$. Lemma 10 says that, assuming the
 uniqueness condition, ${\q'}^1_1$,
 \be
 {\q'}^1_1\equiv\q^1_1-\q^1_1|_{\p^1_{k^1+1}},
 \ee
 is equivalent to $\p^1_{k^1+1}$. Lemma 8 guarantees that
 ${\q'}^1_1={\q'}^1_1|_{\tp_1,\tq_1}$, because the Poisson brackets of
 $\p^1_{k^1_1}$ with $\tp_1$ and $\tq_1$ vanish
 (see redefinition (\ref{red})). Therefore, from lemma 3, ${\q'}^1_1=\op_1{\q'}^1_1$.
 In addition lemma 11 says that ${\q'}^1_1$ is conjugate to $\p^1_1$.
We define $\tp_2\equiv\p^1_1$ and $\tq_2\equiv{\q'}^1_1$.
 Similar to Eq.(\ref{sym1-1}), one can show that
 the constraints,
 \be
 \p^2_{a_2}\in\left\{\p^1_i,\p^1_{k^1+i}|i=1,\cdots,k^1\right\}
 \hspace{1cm}a_2=1,\cdots,2(k-2).
 \ee
 in which we have considered the redefinition,
 \bea
 \p^1_i& \to &\op_2\ \p^1_i,\hspace{1cm}i=1,\cdots,k^1=k-1,\nn\\
 \p^1_{k^1+i}& \to &\op_2\ \p^1_{k^1+i},
 \label{red1}\eea
 where $\op_2=\op_{\tp_2}\op_{\tq_2}$,
 form a set of secondary constraints, i.e.
 \be
 \det\left(\{\p^2_{a_2},\p^2_{b_2}\}\right)_{\cal{M}}\neq 0.
 \ee
 Since $\tp_2=\op_1\tp_2$ and $\tq_2=\op_1\tq_2$, from corollary 5, it can
 be verified that $[\op_1,\op_2]=0$. Therefore,
 using corollary 2, one obtains $\{\p^2_{a_2},\tp_i\}=\{\p^2_{a_2},\tq_i\}=0$,
 $i=1,2$.
 All the above process can be repeated until one ends up with a set of constraints satisfying
 Eq.(\ref{res}).
 \par
 {\bf Lemma 12.} The set of constraints $\tp_i$ and $\tq_i$, $i=1,\cdots,k$, satisfy
 Eq.(\ref{res}).
 \par
 {\bf Proof.} Since
 \bea
 \tp_i&=&\op_j\tp_i,\hspace{1cm}j<i,\nn\\
 \tq_i&=&\op_j\tq_i,
 \eea
 where $\op_i=\op_{\tp_i}\op_{\tq_i}$, corollary 5 reads,
 \be
 [\op_i,\op_j]=0,\hspace{1cm}i,j=1,\cdots,k.
 \label{com1}\ee
 From lemma 3, it can be verified that
 \bea
 \tp_i&=&\tp_i|_{\tp_j,\tq_j}, \hspace{1cm} j<i,\nn\\
 \tq_i&=&\tq_i|_{\tp_j,\tq_j}.
 \eea
 Using lemma 5 one obtains,
 \bea
 \tp_i&=&\tp_i|_{\tp_j,\tq_j}, \hspace{1cm}i\neq j,\nn\\
 \tq_i&=&\tq_i|_{\tp_j,\tq_j}.
 \eea
 Finally, corollary 2 guarantees the validity of lemma 12.
 \par
 When we have found second class constraints satisfying the symplectic algebra, we
 can convert them to first class constraints by extending the phase space to include
 new coordinates $\eta_i$'s and $\pi_i$'s, where
 \bea
 &\{\eta_i,\eta_j\}=\{\pi_i,\pi_j\}=0,&\nn\\
 &\{\pi_i,\eta_j\}=-\delta_{ij},&\nn\\
 &\{\eta_i,z^\mu\}=\{\pi_i,z^\nu\}=0,&
 \eea
 and redefine constraints as follows:
 \bea
 \tp_i&\to&\Phi_i=\tp_i+\eta_i,\nn\\
 \tq_i&\to&\Phi_{k+i}=\tq_i-\pi_i.
 \eea
 It can be easily verified that the constraints $\Phi_a$, $a=1,\cdots,2k$ are abelian,
 \be
 \{\Phi_a,\Phi_b\}=0.
 \ee
 Another interesting result is that, the operator $\op$ defined by the relation,
 \be
 \op\equiv\prod_{i=1}^k\op_i,
 \ee
 is the projection map to the constraint surface $\cal{M}$ and the
 projected coordinates $z^\mu_p\equiv\op z^\mu=z^\mu|_{\cal{M}}$,
 are the coordinates of the constrained surface
 $\cal{M}$.
 In addition, from corollary 2 it is clear that,
 \be
  \{z^\mu_p,\cdots\}_{DB}= \{z^\mu_p,\cdots\}.
 \ee
 where $\{\ ,\ \}_{DB}$ stands for Dirac bracket respective to the constraints
 $\tilde{\p}$'s.
 \par
 Assume one maps each constraint $\tp_i$ to the surface of its conjugate $\tq_i$, i.e.
 \be
 \tp_i\to\tp_i|_{\tq_i},\hspace{1cm}i=1,\cdots,k.
 \label{end1}
 \ee
 From corollary 6, lemma 8 and lemma 12, one verifies that, the algebra (\ref{res}) is
 still satisfied.  Recalling the constraints $\p_a$'s in Eq.(\ref{start}) and the method
 we used to obtain $\tp_i$'s (see  Eqs.(\ref{red},\ref{red1},\ref{end1}) and lemma 10), we verify that
 $\tp_i$'s are simply half of $\p_a$'s, mapped to the surface of $\p_b$'s, $b\neq a$.
 \par
 {\bf Theorem.} Given a set of second class constraints $\p_a$, $a=1,\cdots,2k$,
 where,
 \be
 \p_a=\p_a|_{\p_b},\hspace{1cm}b\neq a,
 \ee
 there exist a permutation $p$ such that the constraints $\tp_{p_i}$, $i=1,\cdots,k$,
 form a subset of abelian (first class) constraints,
 \be
 \{\p_{p_i},\p_{p_j}\}=0,\hspace{1cm}i,j=1,\cdots,k.
 \ee
 As an example see reference \cite{Vyth}, where gauge invariance in the Proca model is
 studied considering the abelian subset of second class constraints.
 \section{Conclusion}
The main purpose of this article is to show that there exist an abelian subset of
second class constraints that can be obtained by mapping each constraint to the
surface of other constraints. In addition we introduced a method that can be
practically used to transform a given set of second class constraints to an
 equivalent set satisfying the symplectic algebra. In this way,
 second class constraints can be simply converted to abelian first class ones.
 \par
 In reference \cite{Abelian}, it is proved that first class constraints become abelian
 when they are mapped to the surface of each other. Therefore one can conclude that,
 using the same technique, the abelian subset of a given set of constraints,
 can be found independent of the details of their algebra.
\par
 Assuming the most general Poisson structure, we have found the projector operators
 that map functions of phase space to the constraint surface.
 It is shown that the Poisson brackets of these mapped functions with other functions
 are equivalent to the corresponding Dirac brackets.
\newpage
\section*{Acknowledgement} The author would like to express his thanks to
  A. Shirzad for valuable discussions and useful comments.


\begin{thebibliography}{99}
\bibitem{Dirac} P. A. M. Dirac, Can. J. Math. {\bf 2}, (1950) 129 ;
Proc. R. Soc. London Ser. A {\bf 246}, (1958) 326; {\it "Lectures on Quantum
Mechanics"} New York: Yeshiva University Press, 1964.
\bibitem{Henbook} M. Henneaux and C. Teitelboim
{\it "Quantization of Gauge System"} Princeton University Press, Princeton, New
Jersey, 1992.
\bibitem{Hen} M. Henneaux, Phys. Rep. {\bf 126}, (1985) 1.
\bibitem{Abelian} F. Loran, Phys. Lett. {\bf B547}, (2002) 63, hep-th/0209180.
\bibitem{Under} I. A. Batalin and E. S. Fradkin, Phys. Lett. {\bf B180}, (1986) 157;\\
I. A. Batalin and I. V. Tyutin, Int. J. Mod. Phys. {\bf A6}, (1991) 3225; \\
I. A. Batalin and R. Marnelius, Mod. Phys. Lett. {\bf A16}, (2001) 1505,
hep-th/0106087.
\bibitem{String} C. S. Chu, P. M. Ho, Nucl.Phys. {\bf B568}, (2000) 447,
hep-th/9906192;\\
 I. Rudychev, JHEP {\bf 0104}, (2001) 015, hep-th/0101039;\\
F. Loran, Phys. Lett. {\bf B544}, (2002) 199, hep-th/0207025.
\bibitem{Covariant}  I. Batalin, S. Lyakhovich and R. Marnelius,
Phys. Lett. {\bf B534}, (2002) 201, hep-th/0112175; \\
 N. Berkovits, JHEP {\bf 0109}. (2001) 016.
\bibitem{Convert} L. Faddeev, Phys. Lett. {\bf B145}, (1984) 81;\\
L. Faddeev and S. Shatashvili, Phys. Lett. {\bf B167}, (1986) 225;\\
E. Sh. Egorian and R. P. Manvelian, Theor. Math. Phys. {\bf 94}, (1993) 173;\\
I. A. Batalin and E. S. Fradkin, Nucl Phys. {\bf B279}, (1987) 514;\\
I. A. Batalin, E. S. Fardkin and T. A. Fradkina, Nucl. Phys. {\bf B314}, (1989) 158.
\bibitem{Vyth} A. S. Vytheeswaran,  Int. J. Mod. Phys. {\bf A13}, (1998) 765,
hep-th/9701050.
\bibitem{John} F. John, "{\it Partial Differential Equations}", vol. 1, Fourth Edition,
Springer-Verlag, New York Inc., 1981.
\end{thebibliography}
\end{document}